\documentclass[a4paper,prd,preprint,tightenlines,floatfix,preprintnumbers,nofootinbib,eqsecnum]{revtex4}

\pdfoutput=1

\usepackage{graphicx}
\usepackage{amssymb, amsmath, amsfonts, bm}

\usepackage{color}

\usepackage[english]{babel}

\usepackage[utf8]{inputenc}

\usepackage[absolute,overlay]{textpos} 

\usepackage{hyperref}

\newcommand{\scaleeight}[1]{%
  {\mathpalette\scaleeightaux{#1}}%
}
\newcommand{\scaleeightaux}[2]{%
  \scalebox{0.85}{$#1#2$}%
}
\newcommand{\subC}[0]{%
  \scaleeight{C}
}
\newcommand{\subD}[0]{%
  \scaleeight{D}
}

\begin{document}


\begin{textblock*}{3cm}(17.7cm,.3cm) 
    \noindent\includegraphics[width=3cm]{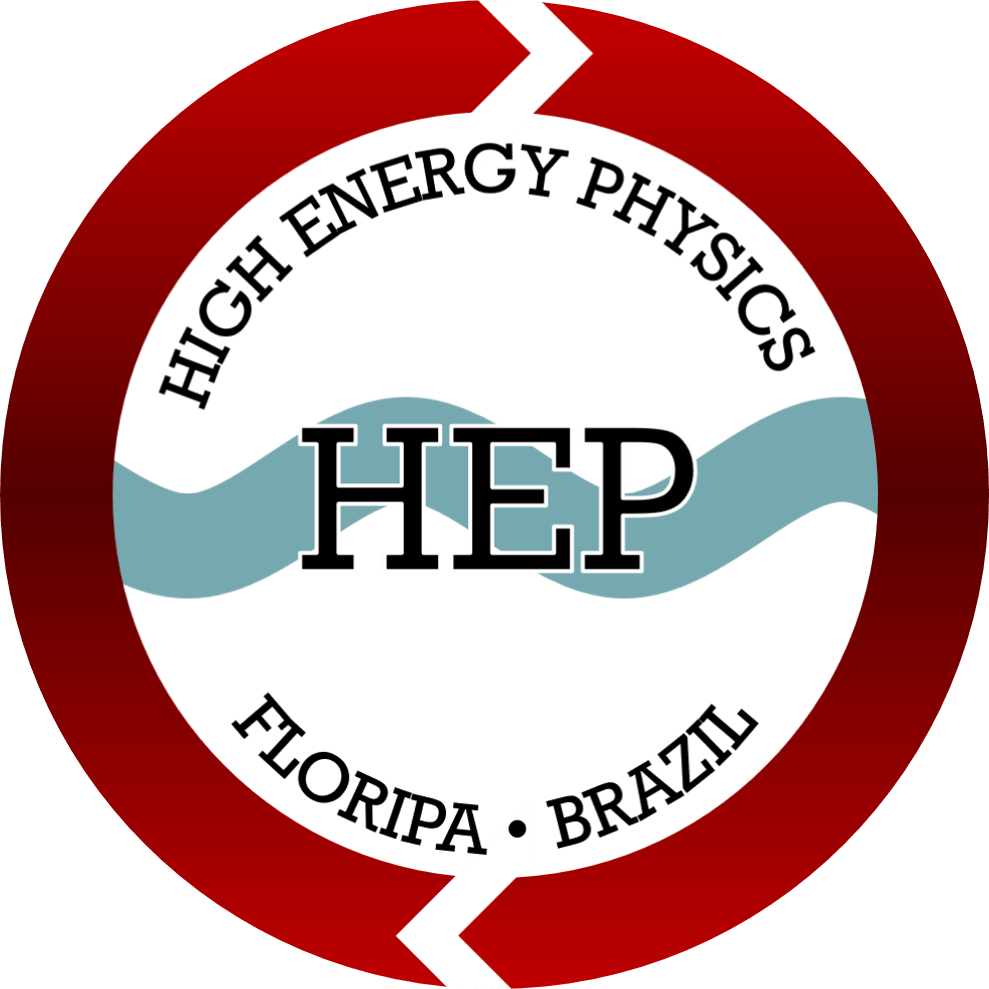} 
\end{textblock*}

\title{
Momentum fraction and hard scale dependence \\ 
of double parton scattering in heavy-ion collisions
\vspace{0.5cm}
}

\author{Joao Vitor C. Lovato}
\email{joaovitorcl1000@gmail.com}

\author{Edgar Huayra}
\email{yuberth022@gmail.com}

\author{Emmanuel G. de Oliveira}
\email{emmanuel.de.oliveira@ufsc.br}

\affiliation{
\\
{$^1$\sl Departamento de F\'isica, CFM, Universidade Federal de Santa Catarina,
C.P. 5064, CEP 88035-972, Florian\'opolis, SC, Brazil}
\vspace{0.7cm}
}

\begin{abstract}
\vspace{0.5cm}

In a previous work, we studied the momentum fraction and hard--scale dependence of double parton scattering (DPS) in proton--proton collisions and the resulting dependence of the effective cross section on the final--state observables. In this paper, we extend those results to heavy--ion ($pA$ and $AA$) collisions, accounting for nuclear effects in the relevant kinematic region, namely shadowing and antishadowing. In addition to modifying the longitudinal parton distributions, these effects also alter the transverse parton distribution of the nucleus, for which we propose a simple new nuclear profile. We further hypothesize that partons inside a bound nucleon are more widely separated than in a free proton. We compute the effective cross section for the available $p$Pb data, obtaining reasonable agreement, and provide predictions for future measurements at the LHC. The observed dependence of our predictions on the final state indicates that DPS in heavy--ion collisions can be used to probe the transverse profile of the free proton and the bound nucleon, primarily in $pA$ collisions, as well as the transverse structure of the nucleus, mainly in $AA$ collisions.

\end{abstract}
\maketitle

\section{Introduction}
\label{Sec:intro}

In high--energy hadron collisions, more than one parton--parton interaction may occur within a single inelastic event, leading to multiple parton interactions (MPI)~\cite{Paver:1982yp, Mekhfi:1983az, Sjostrand:1987su} beyond the standard single--parton scattering (SPS) mechanism. The simplest realization of this phenomenon is double parton scattering (DPS), where two hard interactions take place in the same inelastic event, involving two partons from each of the colliding hadrons, reviewed in Refs.~\cite{Bartalini:2011jp, Bansal:2014paa, Szczurek:2015vha, Diehl:2017wew}. Through two--parton correlations~\cite{Cattaruzza:2005nu, Gaunt:2009re, Snigirev:2010tk, Diehl:2011yj, Blok:2011bu, Chang:2012nw, Blok:2012jr, Blok:2013bpa, Salvini:2013xpz}, DPS processes provide valuable information about the internal structure of the hadron, in particular its less well--known transverse structure~\cite{Strikman:2001gz, dEnterria:2016yhy, dEnterria:2017yhd, Buffing:2017mqm, Gaunt:2018eix, Huayra:2019iun, Huayra:2020iib, Huayra:2021eve, Blok:2022ywz, Huayra:2023gio, Lovato:2025jgh, Stahlhofer:2025tli}.

In the simplified scenario where correlations between the two partons participating in DPS are neglected, one can adopt a purely geometric approach, leading to a fully factorized pocket formula for the DPS cross section:
\begin{align}
\sigma^{\text{DPS}}(CD) = \frac{m}{2} \frac{\sigma^{\text{SPS}}(C)\sigma^{\text{SPS}}(D)}{\sigma_{\text{eff}}}.
\label{eq:nuclear_pocketformula}
\end{align}
Here, $\sigma^{\text{SPS}}(C)$ and $\sigma^{\text{SPS}}(D)$ denote the standard inclusive SPS cross sections for the observables $C$ and $D$, respectively. The quantity $m$ is a symmetry factor ($m = 1$ if $C$ and $D$ are indistinguishable, and $m = 2$ otherwise). The DPS effective cross section $\sigma_{\text{eff}}$ ensures the correct physical dimensions and encodes residual information about the transverse structure of the hadrons. This pocket formula is used for $pp$, $pA$, and $AA$ collisions.

DPS has been extensively measured in proton--proton ($pp$) collisions~\cite{AxialFieldSpectrometer:1986dfj, UA2:1991apc, CDF:1993sbj, CDF:1997yfa, ATLAS:2013aph, D0:2014owy, D0:2014vql, CMS:2013huw, D0:2015rpo, CMS:2015wcf, ATLAS:2016rnd, LHCb:2015wvu, Shao:2016wor, Lansberg:2016muq, ATLAS:2016ydt, LHCb:2016wuo, Lansberg:2017chq, Lansberg:2019adr, CMS:2019jcb, CMS:2021lxi, CMS:2022pio, ALICE:2023lsn, Leontsinis:2022cyi, LHCb:2023qgu, LHCb:2023ybt, CMS:2026evu, ATLAS:2025bcb}. These measurements indicate that $\sigma_{\text{eff}}$ depends on the final--state observables $C$ and $D$. In our previous work~\cite{Lovato:2025jgh}, we developed a phenomenological model that successfully describes this dependence beyond the fully factorized pocket formula, allowing the transverse profile of the proton to depend on the longitudinal momentum fractions of the partons and on the hard energy scales of the subprocesses. More recently, ATLAS measured $\sigma_{\text{eff}}^{\text{exp}} = 10.6 \pm 1.8$\,mb in the 13~TeV $WW$ channel~\cite{ATLAS:2025bcb}. An a posteriori calculation within our framework yields $\sigma_{\text{eff}} = 7.16^{+3.21}_{-3.66}$\,mb, demonstrating the predictive capability of the model.

Recently, there have been two DPS measurements in $p$Pb collisions at the LHC~\cite{CMS:2024wgu, LHCb:2020jse}. Most theoretical studies in the literature split $pA$ DPS into 1x1 and 1x2 contributions~\cite{Strikman:2001gz}, in which the free proton interacts with one or two bound nucleons, respectively. These studies typically employ collinear nuclear PDFs and fixed transverse profiles for free protons, bound nucleons, and nuclei, leading to an effective cross section that is the same for all final states. Motivated by shadowing at small $x$ (reviewed in Ref.~\cite{Klasen:2023uqj}) and Glauber--Gribov modeling (reviewed in Refs.~\cite{Miller:2007ri, dEnterria:2020dwq}), a recent work~\cite{Shao:2020acd} linked DPS measurements to impact--parameter--dependent nPDFs, allowing for different values of the effective cross section but without specifying the transverse profile.

The purpose of this work is to extend our DPS model~\cite{Lovato:2025jgh} to proton--nucleus ($pA$) and nucleus--nucleus ($AA$) collisions, by constructing effective cross sections that depend on the final--state observables $C$ and $D$. To this end, we investigate the dependence of partons inside the nucleus on the longitudinal momentum fraction $x$ and the hard energy scale $\mu$. We 
\begin{enumerate}
    \item take into account that nuclear effects modify the $x$ distribution of partons in a bound nucleon and, consequently, reweight the $x$--dependent transverse profile;
    \item introduce a new (transverse) nuclear profile that phenomenologically incorporates the $x$--dependent shadowing and antishadowing effects without introducing free parameters;
    \item further hypothesize that partons inside a bound nucleon are more widely separated than in a free proton.
\end{enumerate}
This framework leads to agreement with the available experimental measurements. 

We also provide predictions for DPS effective cross sections across a wide range of final states in $p$Pb and PbPb collisions at nucleon--nucleon center--of--mass (c.o.m.) energies $\sqrt{s}=8.16$~TeV and $5.5$~TeV, respectively. Their dependence on the final state indicates that DPS in heavy--ion collisions can be used to probe the transverse profile of the free proton and the bound nucleon, primarily in $pA$ collisions, as well as the transverse structure of the nucleus, mainly in $AA$ collisions.

The paper is organized as follows. In Sec.~II, we detail the theoretical framework for DPS in $pp$, $pA$, and $AA$ collisions, presenting the main ingredients of our approach to calculate the effective cross section in the $pA$ and $AA$ cases. In Sec.~III, we compare our results with available experimental data, provide predictions for channels not yet measured, and discuss their implications for the hadron transverse structure. Finally, in Sec.~IV, we summarize our main findings and interpretations.

\section{Theoretical framework}
\label{Sec:formalism}

In this section, we present the theoretical framework on which our analysis is based. We express the process--dependent effective cross section for the observables $C$ and $D$ as:
\begin{align} \label{eq:nuclear_sigmaeff}
    \frac{1}{\sigma_{\text{eff}}(CD)}
     =& \frac
    {\displaystyle  \int dx_\subC dx_\subD dx_\subC' dx_\subD'\ 
    \Theta(x_\subC, x_\subD; x_\subC', x_\subD'|\, \mu_\subC, \mu_\subD)\ 
    \sigma^C (x_\subC, x_\subC'|\, \mu_\subC)\, 
    \sigma^D (x_\subD, x_\subD'|\, \mu_\subD)}
    {\displaystyle  \int dx_\subC dx_\subC'\, 
     \sigma^C (x_\subC, x_\subC'|\, \mu_\subC)  
      \int dx_\subD dx_\subD'\, 
     \sigma^D (x_\subD, x_\subD'|\, \mu_\subD)} \, .
\end{align}
Here, $x_{\subC,\subD}$ denote the longitudinal momentum fractions of the partons originating in the right--moving hadron (or nucleus), while $x_{\subC,\subD}'$ correspond to those originating in the left--moving one. We take the renormalization and factorization scales to be the hard scales $\mu_\subC$ and $\mu_\subD$ of the processes.

The differential SPS cross section for the observable $C$ is defined as
\begin{align}
    \sigma^{C} (x_\subC, x_\subC'|\,\mu_\subC) \equiv \frac{d \sigma^{\text{SPS}}(C)}{d x_\subC d x'_\subC}
    = \sum_{ik'} f_{i}(x_\subC |\, \mu_\subC)\, 
    \hat{\sigma}_{i k'}^{C} (x_\subC, x_\subC'|\, \mu_\subC)\, 
    f_{k'}(x_\subC' |\, \mu_\subC),
    \label{eq:unintegratedcrosssection}
\end{align}
where $f_i$ and $f_{k'}$ denote the PDFs for parton flavors $i$ and $k'$ and depend on $x_\subC$ or $x'_\subC$, respectively, in addition to the scale $\mu_\subC$. The sum runs over all parton species. An analogous definition holds for the SPS cross section for the observable $D$. We obtain the SPS cross sections using PYTHIA~8.3~\cite{Bierlich:2022pfr}, with the default NNPDF2.3 QCD+QED LO set for the proton~\cite{NNPDF:2017mvq} and the nNNPDF3.0 NLO (Pb--208) set for lead~\cite{AbdulKhalek:2022fyi}. We employ the default PYTHIA parameters, as the PYTHIA default configuration is suitable for broad theoretical calculations that involve non--diffractive, high momentum--transfer collisions at LO. Since $\sigma_{\rm eff}$ is a ratio of cross sections, it should depend only weakly on this choice. We omit PYTHIA MPI and nuclear rescattering effects (e.g.\ as modeled by Angantyr~\cite{Bierlich:2018xfw}). 

So far, no approximation other than working within collinear factorization has been made, as long as the scale factor $\Theta$ is left unspecified. This point is relevant because the information on the number of partons at a given momentum fraction is already absorbed into the PDFs. As a result, the factor $\Theta$ encapsulates all information relevant for DPS coming from the transverse distributions of the two interacting partons, as well as from longitudinal and transverse correlations between them, including:
\begin{itemize}
    \item its dependence on the momentum fractions and the hard scales;
    \item correlations in transverse space;
    \item whether the partons originate from the same or from different nucleons;
    \item corrections to the number of partons when going from SPS to DPS distributions;
    \item the average transverse separation between two partons.
\end{itemize}
On the topic of parton number corrections and sum rules, we refer the reader to the recent Refs.~\cite{Ceccopieri:2025edn,Fedkevych:2025lgp}. In all cases considered in this article, we assume $\Theta$ to be independent of the parton flavor.

\begin{table}
\def\arraystretch{1.4}
    \centering
    \begin{tabular}{|c|c|c|c|c|c|} \hline
    Parameters & $\beta$ (mb)& $\gamma_1$ (mb)& $\gamma_2$ (mb)& $\kappa$ (mb)& $\chi_{\text{dof}}^2$ \\ \hline \hline
    Values & $0.067 \pm 0.068$ & $1.68 \pm 0.48$ & $0.85 \pm 0.16$ & $0.087 \pm 0.036$ & $1.35$ \\ \hline
    \end{tabular}
    \caption{Gaussian variance parameters used in the two--parton transverse profile of free protons, fitted in Ref.~\cite{Lovato:2025jgh}.}
    \label{tab:parametersx}
\end{table}

We summarize our previous work~\cite{Lovato:2025jgh} on proton--proton DPS, as it will be a building block of the approach used here. The transverse distance $r$ between two partons in a proton is described by a normalized Gaussian profile:
\begin{align}\label{eq:F_p}
F_p(x_\subC, x_\subD; r|\, \mu_\subC, \mu_\subD) 
    & = \frac{H(1-x_\subC-x_\subD)}{2\pi B(x_{\subC \subD}, \mu_{\subC \subD})} 
    \exp \left( - \frac{r^2}{2B(x_{\subC \subD}, \mu_{\subC \subD})} \right),
\end{align}
where $x_{\subC \subD} := \sqrt{x_\subC x_\subD}$ and $\mu_{\subC \subD} := \sqrt{\mu_\subC \mu_\subD}$ are geometric means. The kinematic limit $x_\subC + x_\subD \leq 1$ is enforced by the Heaviside step function $H$.

For the Gaussian variance $B(x, \mu)$, we use the following parametrization:
\begin{align}
   B(x, \mu) = \beta + \gamma_1 H(x_0 - x) \ln\!\left(\frac{x_0}{x}\right) 
   + \gamma_2 H(x - x_v) 
   + \kappa \ln\!\left(\frac{\mu}{\mu_0}\right),
   \label{eq:Bwidth}
\end{align}
where $\mu_0 = 1$~GeV, $x_0 = 0.001$, and $x_v = 0.01$. The four free parameters are fitted to experimental data and are given in Table~\ref{tab:parametersx}. At small $x$ and large $\mu$, this parametrization is inspired by Gribov diffusion~\cite{Gribov:1973jg, Brodsky:1994kf}, and related formulae have been discussed in the literature~\cite{Domdey:2009bg, Frankfurt:2010ea, Corke:2011yy}. At large $x$, it follows our previous work on valence or sea quarks~\cite{Huayra:2023gio}.

The scale factor in collisions between two free protons is obtained by integrating the overlap of two Gaussian profiles over $d^2 r$:
\begin{align}\label{eq:theta_pp}
 \Theta_{pp}(x_\subC, x_\subD; x_\subC', x_\subD' |\, \mu_\subC, \mu_\subD) 
  & = \int d^2 r \, 
  F_p(x_\subC, x_\subD; r|\, \mu_\subC, \mu_\subD)\,
  F_p(x_\subC', x_\subD'; r|\, \mu_\subC, \mu_\subD) \nonumber \\
    & = \frac{1}{2\pi} 
    \frac{H(1-x_\subC-x_\subD)\, H(1-x_\subC'-x_\subD')}
    {B(x_{CD}, \mu_{CD}) + B(x_{CD}', \mu_{CD})}.
\end{align}

\begin{figure}
    \centering
    \includegraphics[width=0.7\linewidth]{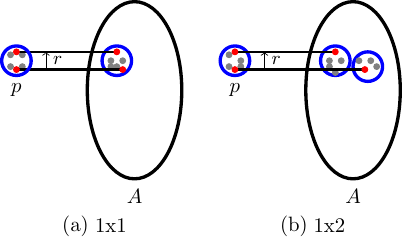}
    \caption{Schematic diagrams of double parton scattering (DPS) contributions in $pA$ collisions, where the two nuclear partons originate either (a) from the same nucleon, labeled 1x1, or (b) from two different nucleons, labeled 1x2.}
    \label{fig:nuclearDPS}
\end{figure}

In $pA$ collisions, we follow Refs.~\cite{Strikman:2001gz, dEnterria:2016yhy, dEnterria:2017yhd} and decompose the scale factor into two contributions:
\begin{eqnarray}\label{eq:theta_pA}
\Theta_{pA} = \frac{\Theta_\text{1x1}}{A} + \frac{(A-1)\Theta_\text{1x2}}{A},
\end{eqnarray}
where $A$ is the nuclear mass number. The label 1x1 indicates that both nuclear partons originate from a single nucleon, while the label 1x2 indicates that they originate from two different nucleons. These contributions are illustrated in Fig.~\ref{fig:nuclearDPS}.

In order to calculate the 1x1 term (also called the DPS1 mechanism in the literature), we do not assume that the parton correlations and transverse distributions in a single nucleon are the same as those of a free proton. In other words, $\Theta_\text{1x1} \neq \Theta_{pp}$. We remark that, even if the 1x1 term were calculated using $\Theta_{pp}$ in Eq.~\ref{eq:nuclear_sigmaeff}, it would not be identical to $1/\sigma_\text{eff}^{pp}$, since it takes into account nuclear rather than proton PDFs and thus weights the Gaussian profiles differently. 

What we actually do is to propose that, at small $x$, two partons inside a nucleon embedded in a nucleus are, on average, more widely separated in transverse space than two partons in a free proton. This is implemented through the following expression:
\begin{eqnarray} \label{eq:theta_1x1}
\Theta_\text{1x1} = \frac{1}{2\pi} 
    \frac{H(1-x_\subC-x_\subD)\, H(1-x_\subC'-x_\subD')}
    {B(x_{CD}, \mu_{CD}) + B(x_{CD}', \mu_{CD}) + \gamma_A H(x_A - x_{CD}') \ln(x_A/x_{CD}')}.
\end{eqnarray}
The values adopted in this work are $\gamma_A = 1$\,mb and $x_A = 5 \times 10^{-3}$. 
At present, the available data are not sufficient to perform a dedicated fit of these parameters, and their choice is therefore justified a posteriori in the Results section. This additional term effectively increases the average transverse distance at small $x$. Its physical origin can be understood via the nuclear strong force, akin to the enlargement of valence electron distributions in a covalent bond. Alternatively, it can represent an effective spreading due to multiple scattering (related to parton recombination) shifting partons that are more widely separated at small $x$ toward larger $x$, or a combination of both effects.

The second term in Eq.~\ref{eq:theta_pA} (also called the DPS2 mechanism), as mentioned above, corresponds to a geometrical coefficient involving nuclear partons originating from different nucleons:
\begin{eqnarray} \label{eq:theta_1x2}
    \Theta_\text{1x2} = \int d^2 r \, 
    F_p(x_\subC, x_\subD; r|\, \mu_\subC, \mu_\subD)\,
    F_A(x_\subC', x_\subD'; r|\, \mu_\subC, \mu_\subD).
\end{eqnarray}
To model this contribution, we assume that partons from different nucleons are uncorrelated:
\begin{eqnarray} \label{eq:double_dist_AA}
    F_A(x'_\subC, x'_\subD; r|\, \mu_\subC, \mu_\subD)
    = \int d^2 r_1\, 
    \rho(x'_\subC; \vec{r}_1|\, \mu_\subC)\, 
    \rho(x'_\subD; \vec{r}_1 + \vec{r}|\, \mu_\subD),
\end{eqnarray}
where $\rho$ denotes the transverse profile of partons inside the nucleus, normalized to unity.

The distribution of nucleons inside the nucleus is well known. We employ the Woods--Saxon~\cite{Woods:1954zz} parametrization:
\begin{eqnarray} \label{eq:WS}
    \rho_\text{WS}(r) = \int_{-\infty}^\infty dz \, 
    \frac{\rho_0}{1 + \exp\!\left[(\sqrt{r^2 + z^2} - R_A)/\delta\right]},
\end{eqnarray}
with $R_A = 6.62$\,fm and $\delta = 0.546$\,fm (from Ref.~\cite{DeVries:1987atn}), and with $\rho_0$ fixed by the normalization condition
\begin{eqnarray} \label{eq:normWS}
    \int d^2 r \, \rho_\text{WS}(r) = 1.
\end{eqnarray}
One could argue that nuclear partons follow the same distribution, as commonly assumed in the literature. Below, we explain why we believe this is not the case.

It is well known that nuclear collinear PDFs are not simply a superposition of free proton and free neutron PDFs. In one approach that we will make use of, the nuclear gluon PDF $f_g^{A}(x|\,\mu)$, for example, can be expressed in terms of the proton gluon PDF $f_g^{p}(x|\,\mu)$ through a nuclear modification factor $R_g(x|\,\mu)$ as
\begin{eqnarray}
    f_g^{A}(x|\,\mu) = R_g(x|\,\mu)\, A\, f_g^{p}(x|\,\mu).
\end{eqnarray}
In the absence of nuclear effects, one has $R_g(x|\,\mu)=1$. At small values of $x$, nuclear effects are mainly driven by multiple scattering. Shadowing occurs for $x \lesssim 0.1$, where $R_g<1$, while antishadowing takes place in the region $0.1 \lesssim x \lesssim 0.3$, where $R_g>1$. In a simplified picture, nuclear partons effectively encode interactions involving more than one parton, redistributing the momentum fractions. As a consequence, relative to free protons, nuclear parton densities are depleted at small $x$ and enhanced at larger values. 


\begin{figure}
    \centering
    \includegraphics[width=0.7\linewidth]{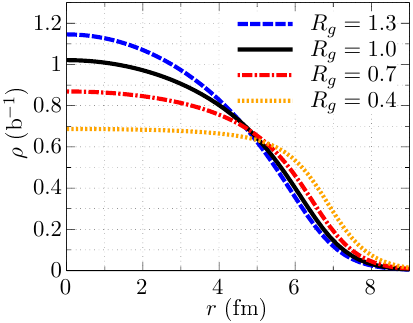}
    \caption{Our new two-dimensional nuclear profile for different values of the nuclear gluon modification factor $R_g$. Compared to the baseline Woods--Saxon distribution at $R_g=1$, larger shadowing (antishadowing) leads to a broadening (narrowing) of the profile. Each curve holds for any combination $(x,\mu)$ giving the indicated $R_g$.}
    \label{fig:rho}
\end{figure}


The next step in our analysis is to construct a longitudinal and transverse distribution of partons inside the nucleus:
\begin{eqnarray} \label{eq:fgAxr}
    f_g^{A}(x; r|\,\mu) = f_g^{A}(x|\,\mu) \rho(x; r|\,\mu).
\end{eqnarray}
Spatial or impact-parameter dependence of nuclear modifications has been studied using, among other approaches, leading--twist nuclear shadowing, spatially dependent nuclear PDFs, and the color--dipole formalism~\cite{Frankfurt:2011cs,Helenius:2012wd,Kopeliovich:2022jwe}.

In order to incorporate nuclear effects generated by multiple scatterings (shadowing and antishadowing) into the transverse distribution of partons, we propose the following new nuclear profile:
\begin{eqnarray} \label{eq:rhoexp}
    \rho(x; r|\,\mu) = \frac{\exp \!\left(\Delta \sigma \, \rho_\text{WS}(r)\right) - 1}{\Delta \sigma \, R_g(x|\,\mu)}.
\end{eqnarray}
The area parameter $\Delta \sigma$ is fixed uniquely for each value of $R_g$ by enforcing the normalization condition
\begin{eqnarray} \label{eq:norm_rho}
    \int d^2 r \, \rho(x; r | \mu) = 1.
\end{eqnarray}
Thus, this framework contains no free parameters. The sign of $\Delta \sigma$ is determined by $R_g$: it is positive when $R_g>1$, negative when $R_g<1$, and vanishes for $R_g=1$. At large $r$, in the dilute regime, the profile reduces to the standard $\rho_\text{WS}$, up to an overall normalization factor.

The gluon nuclear modification factor is used as a proxy for the modification of the transverse distribution of all parton species. This choice is motivated by the fact that gluons and sea quarks dominate the SPS cross sections considered in this work. As illustrated in Fig.~\ref{fig:rho}, the nuclear profile broadens when $R_g < 1$, as effective parton recombination is more pronounced in the nuclear core. In contrast, for $R_g>1$ the profile becomes more compact, which in turn enhances the DPS event rate. We use the $R_g$--dependent nuclear profile only for $x < 0.3$. For larger values of $x$, nuclear effects are not dominated by multiple scattering, and we instead use the Woods--Saxon profile. For small values of $\mu$, the provided values of $R_g$ can become very small; therefore, we impose a lower bound $R_g \ge 0.1$. The nuclear modification factor is obtained from the nNNPDF3.0 NLO sets for Pb and $p$~\cite{AbdulKhalek:2022fyi}.

In $AA$ collisions, the scale factor is decomposed into four contributions:
\begin{eqnarray}\label{eq:theta_AA}
\Theta_{AA} = \frac{\Theta_\text{1x1}}{A^2} + \frac{(A-1)\,\Theta_\text{1x2}}{A^2}
+ \frac{(A-1)\,\Theta_\text{2x1}}{A^2} + \frac{(A-1)^2\,\Theta_\text{2x2}}{A^2},
\end{eqnarray}
where the labels 1 and 2 denote the number of participant nucleons in each nucleus. The first two contributions are analogous to those in $pA$ collisions, with the replacement
\begin{eqnarray}
B(x_{CD}, \mu_{CD}) \rightarrow B(x_{CD}, \mu_{CD}) + \gamma_A H(x_A - x_{CD}) \ln(x_A/x_{CD}),
\end{eqnarray}
in Eqs.~\ref{eq:F_p} and \ref{eq:theta_1x1}, due to the projectile being a bound nucleon. The new 2x1 contribution is symmetric to the 1x2 one, and the new 2x2 contribution is given by
\begin{eqnarray} \label{eq:theta_2x2}
    \Theta_\text{2x2} = \int d^2 r \, 
    F_A(x_\subC, x_\subD; r|\, \mu_\subC, \mu_\subD)\, 
    F_A(x_\subC', x_\subD'; r|\, \mu_\subC, \mu_\subD),
\end{eqnarray}
with the function $F_A$ defined in Eq.~\ref{eq:double_dist_AA}.
The effective cross section in $AA$ collisions is calculated by using $\Theta_{AA}$ in Eq.~\ref{eq:nuclear_sigmaeff}. The result is two orders of magnitude larger than in $pp$ collisions because the participating partons are roughly one order of magnitude farther apart in the transverse plane.

In this formulation, $\sigma_{\text{eff}}(CD)$ is not a universal quantity across different observables $C$ and $D$, as each SPS cross section weights the corresponding scale factor differently. Consequently, by measuring $\sigma_{\text{eff}}(CD)$ for various final states, one can infer the $x$--dependence of the scale factor and thereby extract information about the transverse parton distributions inside hadrons.

\section{Results and predictions}
\label{Sec:prediction}


\begin{figure}
    \centering
    \includegraphics[width=\linewidth]{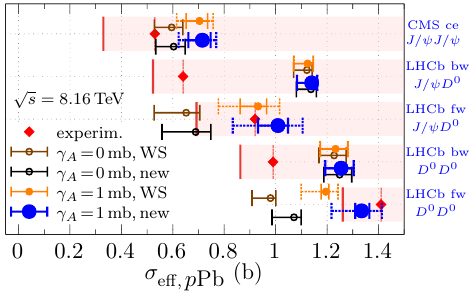}
    \caption{Comparison between available experimental lower limits (in red) and theoretical predictions for the DPS effective cross section in $p$Pb collisions at $\sqrt{s}=8.16$~TeV for various final states $CD$. The CMS result corresponds to the central rapidity (``ce'') double $J/\psi$ result~\cite{CMS:2024wgu}. The LHCb results correspond to $J/\psi + D^0$ and double $D^0$ production in the forward (``fw'') and backward (``bw'') rapidity configurations~\cite{LHCb:2020jse}. Theoretical predictions are shown for bound nucleons with transverse profiles the same as ($\gamma_A=0$, black empty circles) or wider than ($\gamma_A=1$\,mb, blue filled circles) those of free protons, with the new nuclear profile. Predictions with the standard Woods--Saxon profile are also shown ($\gamma_A=0$, brown empty circles; $\gamma_A=1$\,mb, orange filled circles). Solid vertical bars represent the combined $1\sigma$ theoretical uncertainties, while dashed bars indicate the additional variation obtained by varying $\gamma_A$ in the range $0.5$--$1.5$\,mb.}
    \label{fig:exp-pA}
\end{figure}


Having developed the formalism to calculate DPS observables in heavy--ion collisions, we now turn to the experimental data. Measurements of the DPS effective cross section have the advantage that several sources of uncertainty partially cancel. In Fig.~\ref{fig:exp-pA}, we show the five available experimental data points for $p$Pb collisions, namely the CMS 8.16~TeV double $J/\psi$ measurement~\cite{CMS:2024wgu} and the LHCb 8.16~TeV results for $J/\psi$ and $D^0$ combinations~\cite{LHCb:2020jse}. In all five cases, only lower limits are available, since the SPS contribution to the same final states is not precisely estimated or fully subtracted. The nucleon--nucleon c.o.m.\ rapidity region is central, $|y|<2.4$, for CMS, while the LHCb regions correspond to backward, $-5.0<y<-2.5$, and forward, $1.5<y<4.0$, rapidities.

In the same Fig.~\ref{fig:exp-pA}, we present four sets of $pA$ predictions employing either the standard Woods--Saxon profile (labeled ``WS'') or the new transverse nuclear profile of Eq.~\ref{eq:rhoexp} (labeled ``new'') and using either $\gamma_A = 0$\,mb (empty circles) or $\gamma_A = 1$\,mb (filled circles) in Eq.~\ref{eq:theta_1x1}. These choices of $\gamma_A$ assume that the transverse distance between two partons inside the same nucleon is the same as, or larger than, the one in a free proton, respectively. We adopt the same experimental cuts as in the corresponding measurements.

We verified in our calculations that the 1x2 contribution to the DPS cross section is of the same order of magnitude as the 1x1 contribution. Shadowing modifies the new nuclear profile shown in Fig.~\ref{fig:rho}, spreading the partons in the 1x2 contribution; however, this effect is shown to be small in Fig.~\ref{fig:exp-pA}. Therefore, the dependence on the final state, namely on the produced particle species and the rapidity, is driven by the 1x1 contribution, on which we will focus next.

For $\gamma_A = 0$\,mb, the CMS central--rapidity $\sigma_\text{eff}$ is the smallest (corresponding to a larger DPS rate), since partons in the region $0.001 < x < 0.01$ are closer in transverse space, as can be seen from Eq.~\ref{eq:Bwidth}. The backward--rapidity values of $\sigma_{\text{eff},\,p\textrm{Pb}}$ are significantly larger than the forward ones for the same observable. This occurs because shadowing in the nuclear PDFs entering Eq.~\ref{eq:nuclear_sigmaeff} is more relevant in the latter case, where the typical momentum fractions in the nucleus are smaller. This depletes the gluon density at small $x$, where partons in the 1x1 contribution are more widely separated, and therefore decreases the effective cross section. The trend observed in the LHCb data goes in the opposite direction to this set of predictions.

This failure constitutes our main motivation to hypothesize that partons inside a nucleon are more spread out than in a free proton. Consequently, the additional $\gamma_A = 1$\,mb term introduced at small $x$ in Eq.~\ref{eq:theta_1x1} leaves the backward results essentially unchanged while increasing the forward effective cross section. This set of predictions remains compatible with the experimental results within the current uncertainties, given that the experimental upper error bars have not been determined. Future measurements in $pA$ collisions may help constrain the value of $\gamma_A$. In addition, they may provide information on the transverse separation between partons inside the free proton.

Our solid uncertainty bars reflect the two most important sources of theoretical uncertainty associated with the DPS effective cross section for a fixed value of $\gamma_A$, and correspond to one--standard--deviation ($1\sigma$) variations. The first source arises from the $\Theta_{\textrm{1x1}}$ model established in our previous work on $pp$ collisions~\cite{Lovato:2025jgh}. To estimate this uncertainty, we use the Hessian matrix from the original Minuit2~\cite{James:2004xla} fit and generate 200 replicas. The second source originates from the nuclear modification factor $R_g$ entering $\Theta_{\textrm{1x2}}$, for which we employ the 200 replicas provided by the nNNPDF3.0 NLO set. The EPPS21~\cite{Eskola:2021nhw} distribution was also tested and showed agreement within the error bars.

In addition, for our main $\gamma_A = 1$\,mb set of predictions, we provide extra dashed bars exploring the range $\gamma_A = 0.5$--$1.5$\,mb, combined with the previously discussed uncertainties, resulting in a wider bar in Fig.~\ref{fig:exp-pA}. This variation does not represent a statistical uncertainty, but rather provides an estimate of the theoretical uncertainty associated with the modeling of the 1x1 contribution. Variations of the parameter $x_A$ lead to correlated effects and can be partially compensated by corresponding changes in $\gamma_A$.

\begin{figure}[t]
    \centering
    \includegraphics[width=\linewidth]{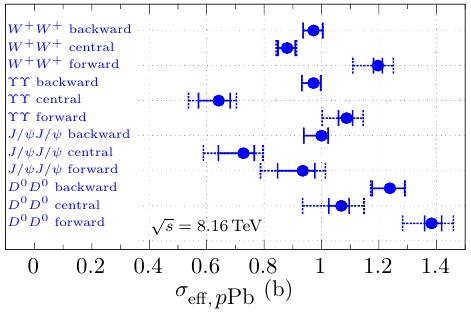}\hfill
    \caption{Theoretical predictions for the DPS effective cross section in $\sqrt{s} = 8.16$~TeV $p$Pb collisions for final states with identical observables ($C=D$), considering backward, central, and forward rapidity configurations.}
    \label{fig:pred-pA-XX}
\end{figure}

\begin{figure}[t]
    \centering
    \includegraphics[width=\linewidth]{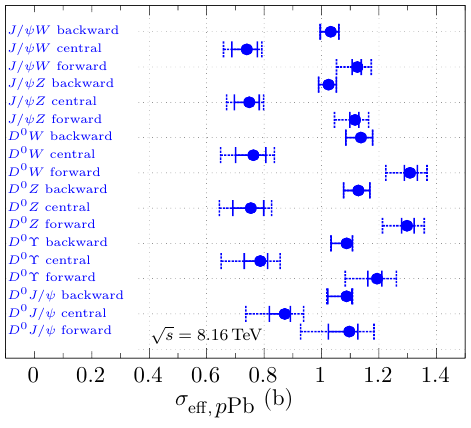}
    \caption{Theoretical predictions for the DPS effective cross section in $\sqrt{s} = 8.16$~TeV $p$Pb collisions for final states with different observables ($C\neq D$), considering backward, central, and forward rapidity configurations.}
    \label{fig:pred-pA-XY}
\end{figure}

\begin{figure}
    \centering
    \includegraphics[width=\linewidth]{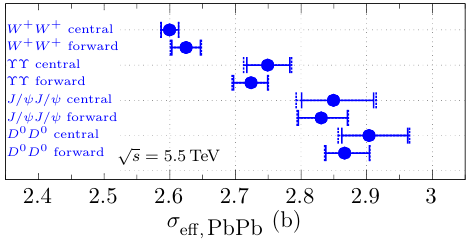}\hfill
    \caption{Theoretical predictions for the DPS effective cross section in $\sqrt{s} = 5.5$~TeV PbPb collisions for final states with identical observables ($C=D$), considering central and forward rapidity configurations.
    }
    \label{fig:pred-AA-XX}
\end{figure}

\begin{figure}
    \centering
    \includegraphics[width=\linewidth]{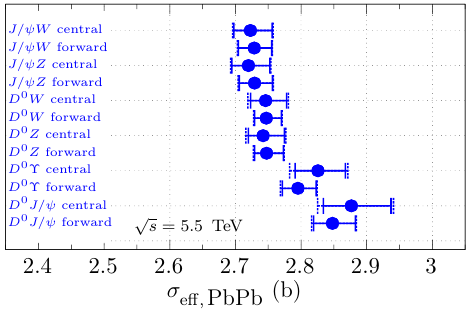}
    \caption{Theoretical predictions for the DPS effective cross section in $\sqrt{s} = 5.5$~TeV PbPb collisions for final states with different observables ($C\neq D$), considering central and forward rapidity configurations.}
    \label{fig:pred-AA-XY}
\end{figure}


We now move on to predictions for final states that have not yet been observed and do not pursue further results with $\gamma_A=0$\,mb. In Figs.~\ref{fig:pred-pA-XX} and~\ref{fig:pred-pA-XY}, we present our predictions for final states with identical ($C=D$) and different ($C\ne D$) observables in 8.16~TeV $p$Pb collisions. Three rapidity ranges are considered: backward ($-4.5<y<-2.0$), central ($|y|<2.0$), and forward ($2.0<y<4.5$). No additional cuts are introduced with the exception of a minimum $W$ transverse mass requirement in the corresponding SPS simulations~\cite{CMS:2013huw}.

For the same final state $CD$, the rapidity hierarchy $\sigma_\text{eff}^\text{central} \ll \sigma_\text{eff}^\text{backward} \lesssim \sigma_\text{eff}^\text{forward}$ of Fig.~\ref{fig:exp-pA} is observed again in the predictions. This behavior admits the same explanation as discussed previously. Final states involving quarkonia, such as $J/\psi$ and $\Upsilon$ pairs, exhibit a larger DPS contribution than processes such as double $D^0$ meson or double same--sign $W$ production. This pattern was already observed in $pp$ collisions~\cite{Lovato:2025jgh}.

We also present predictions for 5.5~TeV PbPb collisions at central ($|y|<2$) and forward (noncentral, $2<|y|<4.5$) rapidities, as shown in Figs.~\ref{fig:pred-AA-XX} and~\ref{fig:pred-AA-XY}. In this case, the $\Theta_{2x2}$ term is about two orders of magnitude larger than the others. Therefore, in contrast to $pA$ collisions, the new nuclear profile influenced by nuclear PDF effects of Eq.~\ref{eq:rhoexp} plays the dominant role. If the transverse profile is instead fixed to the standard Woods--Saxon form, the $\Theta_{2x2}$ contribution yields $\sigma_{\text{eff},\,AA}\simeq2.7$~b independent of the final state, up to the small subleading corrections. We remark that the uncertainty associated with $\gamma_A$ has a much smaller impact, as the dashed and continuous error bars are almost coincident in the figures.

Comparing different final states within the same rapidity range, we find that final states composed of lighter observables $C$ and $D$ lead to a larger effective cross section. These processes typically probe smaller longitudinal momentum fractions $x$ and are therefore more sensitive to shadowing effects. Heavier observables probe larger values of $x$, where shadowing is reduced, or antishadowing may occur. Shadowing broadens the nuclear profile of Fig.~\ref{fig:rho}, while antishadowing narrows it. This behavior explains the observed variation of the effective cross section across different final states.

Considering the same final state $CD$ at different rapidities, two qualitatively different behaviors can arise. At central rapidities, partons in both nuclei typically probe similar values of $x$ and therefore experience comparable nuclear modification factors $R_g$. Moving to forward rapidities, partons in the right--moving nucleus probe larger values of $x$ (and thus larger $R_g$), while partons in the left--moving nucleus probe smaller values of $x$ (and smaller $R_g$). If the typical central--rapidity kinematics lies in the antishadowing region, the transition to forward rapidities enhances shadowing in the left--moving nucleus while leaving the amount of antishadowing in the right--moving nucleus essentially unchanged. The net effect is an increase of the effective cross section, as the widening of the left--moving profile dominates over the narrowing of the right--moving one. This behavior is observed, for instance, in double $W^+$ production.

For lighter final states such as double $D^0$ or $J/\psi$ production, the situation is reversed. In this case, central rapidities already probe the shadowing region. When moving to forward rapidities, the narrowing of the right--moving nuclear profile dominates over the widening of the left--moving one, leading to a decrease of the effective cross section.

Overall, our results show that DPS effective cross sections in nuclear collisions are sensitive to different dynamical mechanisms, depending on the colliding system and kinematic regime. In $pA$ collisions, the observable and rapidity dependence is primarily driven by the 1x1 contribution, making these measurements particularly sensitive to the transverse parton distribution inside a bound nucleon. In contrast, in $AA$ collisions, the DPS cross section is dominated by the 2x2 contribution, where shadowing and antishadowing nuclear effects determine the transverse profile. This complementarity highlights the potential of DPS measurements in nuclear collisions to probe free proton, bound nucleon, and nuclear structures, motivating future experimental studies.

\section{Conclusions}
\label{Sec:conclusions}

In this paper, we have studied double parton scattering (DPS) in proton--nucleus ($pA$) and nucleus--nucleus ($AA$) collisions, using Pb as the example nucleus. Before integrating over the parton momentum fractions, we factorize the DPS cross section (Eq.~\ref{eq:nuclear_sigmaeff}) as the product of two single parton scattering (SPS) cross sections and a scale factor $\Theta$ that depends on the momentum fractions and the hard scales of the subprocesses. In contrast to the standard pocket formula, our effective cross sections $\sigma_{\text{eff},\,pA}$ and $\sigma_{\text{eff},\,AA}$ take different numerical values depending on the chosen observables $C$ and $D$, i.e., on the produced particles and their rapidity ranges (and possibly other kinematic cuts).

The scale factor, and thus the effective cross section, is sensitive to the transverse geometry (profile) of the colliding particles and depends on the parton distributions and their correlations. The profile is a function of the transverse distance between the two partons inside the hadron. As the profile becomes wider, the effective cross section increases, corresponding to a smaller DPS rate.

For the free proton, we used the Gaussian profile from our previous work, whose $x$-- and $\mu$--dependent width was fitted to $pp$ DPS data~\cite{Lovato:2025jgh}. The difference here is that this profile is convolved with nuclear PDFs. In this paper, we proposed that partons inside the same bound nucleon are more separated than in free protons at small $x$. We also proposed a new nuclear profile characterizing partons from different nucleons modified by shadowing and antishadowing effects, which are mainly driven by parton recombination. Shadowing (antishadowing) decreases (increases) the normalized transverse density at the center of the profile, making it wider (narrower).

In $pA$ collisions, there are two possible interaction mechanisms: 1x1, in which both nuclear partons originate from a single nucleon, and 1x2, in which they originate from two different nucleons. The 1x1 contribution, although about the same size as the 1x2 one, exhibits a larger absolute variation and is therefore the dominant contribution controlling the variation of the effective cross section. Depending on the observable and rapidity, $\sigma_{\text{eff},\,pA}$ can vary by about 50\%. We obtain a good description of the available data and provide predictions for future measurements. This agreement supports the hypothesis of more widely separated partons inside the same nucleon. 

In $AA$ collisions, four contributions arise: 1x1, 1x2, 2x1, and 2x2, where the labels 1 and 2 denote the number of participant nucleons in each nucleus. The first three contributions are analogous to those in the $pA$ case. The new 2x2 contribution is dominant, being about two orders of magnitude larger than the others, and depends only on shadowing and antishadowing effects, not on correlations inside the nucleon. We provide predictions for the effective cross section $\sigma_{\text{eff},\,AA}$, which has not yet been measured. By varying the final--state observables, the numerical value of $\sigma_{\text{eff},\,AA}$ can change by about 10\%.

Overall, our results predict a non--trivial variation of $\sigma_{\text{eff},\,pA}$ and $\sigma_{\text{eff},\,AA}$ among different final states. In the $pA$ case, this offers an opportunity to study correlations inside the nucleon, while in the $AA$ case, it provides a probe of how shadowing and antishadowing modify the nuclear transverse profile. Our predictions can be tested in future measurements, and we emphasize the importance of complementary channels to further constrain double parton scattering in heavy--ion collisions.

\section*{Acknowledgments}

This work was supported by FAPESC, INCT-FNA, and CNPq (Brazil).



\begin{thebibliography}{10}

\bibitem{Paver:1982yp}
N.~Paver and D.~Treleani,
Nuovo Cim. A \textbf{70}, 215 (1982).

\bibitem{Mekhfi:1983az}
M.~Mekhfi,
Phys. Rev. D \textbf{32}, 2371 (1985).

\bibitem{Sjostrand:1987su}
T.~Sjostrand and M.~van Zijl,
Phys. Rev. D \textbf{36}, 2019 (1987).

\bibitem{Bartalini:2011jp}
P.~Bartalini \textit{et al.}
[arXiv:1111.0469 [hep-ph]].

\bibitem{Bansal:2014paa}
S.~Bansal \textit{et al.}
[arXiv:1410.6664 [hep-ph]].

\bibitem{Szczurek:2015vha}
A.~Szczurek,
Acta Phys. Polon. Supp. \textbf{8}, no.2, 483 (2015)
[arXiv:1505.04067 [hep-ph]].

\bibitem{Diehl:2017wew}
M.~Diehl and J.~R.~Gaunt,
Adv. Ser. Direct. High Energy Phys. \textbf{29}, 07 (2018)
[arXiv:1710.04408 [hep-ph]].

\bibitem{Cattaruzza:2005nu}
E.~Cattaruzza, A.~Del Fabbro, and D.~Treleani,
Phys. Rev. D \textbf{72}, 034022 (2005)
[arXiv:hep-ph/0507052 [hep-ph]].

\bibitem{Gaunt:2009re}
J.~R.~Gaunt and W.~J.~Stirling,
JHEP \textbf{03}, 005 (2010)
[arXiv:0910.4347 [hep-ph]].

\bibitem{Snigirev:2010tk}
A.~M.~Snigirev,
Phys. Rev. D \textbf{81}, 065014 (2010)
[arXiv:1001.0104 [hep-ph]].

\bibitem{Diehl:2011yj}
M.~Diehl, D.~Ostermeier, and A.~Schafer,
JHEP \textbf{03}, 089 (2012)
[arXiv:1111.0910 [hep-ph]].

\bibitem{Blok:2011bu}
B.~Blok, Y.~Dokshitser, L.~Frankfurt, and M.~Strikman,
Eur. Phys. J. C \textbf{72}, 1963 (2012)
[arXiv:1106.5533 [hep-ph]].

\bibitem{Chang:2012nw}
H.~M.~Chang, A.~V.~Manohar, and W.~J.~Waalewijn,
Phys. Rev. D \textbf{87}, 034009 (2013)
[arXiv:1211.3132 [hep-ph]].

\bibitem{Blok:2012jr}
B.~Blok, M.~Strikman, and U.~A.~Wiedemann,
Eur. Phys. J. C \textbf{73}, 
2433 (2013)
[arXiv:1210.1477 [hep-ph]].

\bibitem{Blok:2013bpa}
B.~Blok, Y.~Dokshitzer, L.~Frankfurt, and M.~Strikman,
Eur. Phys. J. C \textbf{74}, 2926 (2014)
[arXiv:1306.3763 [hep-ph]].

\bibitem{Salvini:2013xpz}
S.~Salvini, D.~Treleani, and G.~Calucci,
Phys. Rev. D \textbf{89}, 
016020 (2014)
[arXiv:1309.6201 [hep-ph]].

\bibitem{Strikman:2001gz}
M.~Strikman and D.~Treleani,
Phys. Rev. Lett. \textbf{88}, 031801 (2002)
[arXiv:hep-ph/0111468 [hep-ph]].

\bibitem{dEnterria:2016yhy}
D.~d'Enterria and A.~M.~Snigirev,
Eur. Phys. J. C \textbf{78}, no.5, 359 (2018)
[arXiv:1612.08112 [hep-ph]].

\bibitem{dEnterria:2017yhd}
D.~d'Enterria and A.~M.~Snigirev,
Adv. Ser. Direct. High Energy Phys. \textbf{29}, 159-187 (2018)
[arXiv:1708.07519 [hep-ph]].

\bibitem{Buffing:2017mqm}
M.~G.~A.~Buffing, M.~Diehl, and T.~Kasemets,
JHEP \textbf{01}, 044 (2018)
[arXiv:1708.03528 [hep-ph]].

\bibitem{Gaunt:2018eix}
J.~R.~Gaunt and T.~Kasemets,
Adv. High Energy Phys. \textbf{2019}, 3797394 (2019)
[arXiv:1812.09099 [hep-ph]].

\bibitem{Huayra:2019iun}
E.~Huayra, E.~G.~de Oliveira, and R.~Pasechnik,
Eur. Phys. J. C \textbf{79}, 
880 (2019)
[arXiv:1905.03294 [hep-ph]].

\bibitem{Huayra:2020iib}
E.~Huayra, E.~G.~de Oliveira, and R.~Pasechnik,
Eur. Phys. J. C \textbf{80}, 
772 (2020)
[arXiv:2003.06412 [hep-ph]].

\bibitem{Huayra:2021eve}
E.~Huayra, E.~G.~de Oliveira, R.~Pasechnik, and B.~O.~Stahlh{\"o}fer,
Phys. Rev. D \textbf{104}, 
096003 (2021)
[arXiv:2106.11465 [hep-ph]].

\bibitem{Blok:2022ywz}   
B.~Blok, R.~Segev, and M.~Strikman,
Eur. Phys. J. C \textbf{83}, 
415 (2023)
[arXiv:2212.08848 [hep-ph]].

\bibitem{Huayra:2023gio}
E.~Huayra, J.~V.~C.~Lovato, and E.~G.~de Oliveira,
JHEP \textbf{09}, 177 (2023)
[arXiv:2305.11106 [hep-ph]].

\bibitem{Lovato:2025jgh}
J.~V.~C.~Lovato, E.~Huayra, and E.~G.~de Oliveira,
JHEP \textbf{10}, 163 (2025)
[arXiv:2506.05337 [hep-ph]].

\bibitem{Stahlhofer:2025tli}
B.~O.~Stahlh{\"o}fer, E.~Huayra, and E.~G.~de Oliveira,
Phys. Rev. D \textbf{113}, 094015 (2026)
[arXiv:2510.05420 [hep-ph]].

\bibitem{AxialFieldSpectrometer:1986dfj}
T.~\r{A}kesson \textit{et al.} (AFS Collaboration),
Z. Phys. C \textbf{34}, 163 (1987).

\bibitem{UA2:1991apc}
J.~Alitti \textit{et al.} [UA2],
Phys. Lett. B \textbf{268}, 145-154 (1991).

\bibitem{CDF:1993sbj} 
F.~Abe \textit{et al.} (CDF Collaboration),
Phys. Rev. D \textbf{47}, 4857 (1993).

\bibitem{CDF:1997yfa}
F.~Abe \textit{et al.} (CDF Collaboration),
Phys. Rev. D \textbf{56}, 3811 (1997).

\bibitem{ATLAS:2013aph}
G.~Aad \textit{et al.} (ATLAS Collaboration),
New J. Phys. \textbf{15}, 033038 (2013)
[arXiv:1301.6872 [hep-ex]].

\bibitem{D0:2014owy} 
V.~M.~Abazov \textit{et al.} (D0 Collaboration),
Phys. Rev. D \textbf{89}, 072006 (2014)
[arXiv:1402.1550 [hep-ex]].

\bibitem{D0:2014vql} 
V.~M.~Abazov \textit{et al.} (D0 Collaboration),
Phys. Rev. D \textbf{90}, 111101 (2014)
[arXiv:1406.2380 [hep-ex]].

\bibitem{CMS:2013huw} 
S.~Chatrchyan \textit{et al.} (CMS Collaboration),
JHEP \textbf{03}, 032 (2014)
[arXiv:1312.5729 [hep-ex]].

\bibitem{D0:2015rpo}
V.~M.~Abazov \textit{et al.} (D0 Collaboration),
Phys. Rev. D \textbf{93}, 052008 (2016)
[arXiv:1512.05291 [hep-ex]].

\bibitem{CMS:2015wcf} 
V.~Khachatryan \textit{et al.} (CMS Collaboration),
Eur. Phys. J. C \textbf{76}, 155 (2016)
[arXiv:1512.00815 [hep-ex]].

\bibitem{ATLAS:2016rnd} 
M.~Aaboud \textit{et al.} (ATLAS Collaboration), 
JHEP \textbf{11}, 110 (2016)
[arXiv:1608.01857 [hep-ex]].

\bibitem{LHCb:2015wvu} 
R.~Aaij \textit{et al.} (LHCb Collaboration),
JHEP \textbf{07}, 052 (2016)
[arXiv:1510.05949 [hep-ex]].

\bibitem{Shao:2016wor}
H.~S.~Shao and Y.~J.~Zhang,
Phys. Rev. Lett. \textbf{117}, 062001 (2016)
[arXiv:1605.03061 [hep-ph]].

\bibitem{Lansberg:2016muq}
J.~P.~Lansberg and H.~S.~Shao,
Nucl. Phys. B \textbf{916}, 132 (2017). 
[arXiv:1611.09303 [hep-ph]].

\bibitem{ATLAS:2016ydt} 
M.~Aaboud \textit{et al.} (ATLAS Collaboration),
Eur. Phys. J. C \textbf{77}, 76 (2017)
[arXiv:1612.02950 [hep-ex]].

\bibitem{LHCb:2016wuo} 
R.~Aaij \textit{et al.} (LHCb Collaboration), 
JHEP \textbf{06}, 047 (2017)
[arXiv:1612.07451 [hep-ex]].

\bibitem{Lansberg:2017chq} 
J.~P.~Lansberg, H.~S.~Shao, and N.~Yamanaka,
Phys. Lett. B \textbf{781}, 485 (2018)
[arXiv:1707.04350 [hep-ph]].

\bibitem{Lansberg:2019adr} 
J.~P.~Lansberg,
Phys. Rept. \textbf{889}, 01 (2020)
[arXiv:1903.09185 [hep-ph]].

\bibitem{CMS:2019jcb} 
A.~M.~Sirunyan \textit{et al.} (CMS Collaboration),
Eur. Phys. J. C \textbf{80}, 41 (2020)
[arXiv:1909.06265 [hep-ex]].

\bibitem{CMS:2021lxi}  
A.~Tumasyan \textit{et al.} (CMS Collaboration),
JHEP \textbf{01}, 177 (2022)
[arXiv:2109.13822 [hep-ex]].

\bibitem{CMS:2022pio} 
A.~Tumasyan \textit{et al.} (CMS Collaboration),
Phys. Rev. Lett. \textbf{131}, 091803 (2023)
[arXiv:2206.02681 [hep-ex]].

\bibitem{ALICE:2023lsn} 
S.~Acharya \textit{et al.} (ALICE Collaboration),
Phys. Rev. C \textbf{108}, 045203 (2023)
[arXiv:2303.13431 [hep-ex]].

\bibitem{Leontsinis:2022cyi} 
A. Tumasyan \textit{et al.} (CMS Collaboration),
Nature Phys. \textbf{19}, 338 (2023)
[arXiv:2111.05370 [hep-ex]].

\bibitem{LHCb:2023qgu} 
R.~Aaij \textit{et al.} (LHCb Collaboration),
JHEP \textbf{08}, 093 (2023)
[arXiv:2305.15580 [hep-ex]].

\bibitem{LHCb:2023ybt} 
R.~Aaij \textit{et al.} (LHCb Collaboration),
JHEP \textbf{03}, 088 (2024)
[arXiv:2311.14085 [hep-ex]].

\bibitem{CMS:2026evu}
A.~Hayrapetyan \textit{et al.} [CMS],
[arXiv:2602.02770 [hep-ex]].

\bibitem{ATLAS:2025bcb}
G.~Aad \textit{et al.} [ATLAS],
Phys. Lett. B \textbf{870}, 139892 (2025)
[arXiv:2505.08313 [hep-ex]].

\bibitem{CMS:2024wgu}
A.~Hayrapetyan \textit{et al.} [CMS],
Phys. Rev. D \textbf{110}, 
092002 (2024)
[arXiv:2407.03223 [nucl-ex]].

\bibitem{LHCb:2020jse}
R.~Aaij \textit{et al.} [LHCb],
Phys. Rev. Lett. \textbf{125}, 
212001 (2020)
[arXiv:2007.06945 [hep-ex]].

\bibitem{Klasen:2023uqj}
M.~Klasen and H.~Paukkunen,
Ann. Rev. Nucl. Part. Sci. \textbf{74}, 49-87 (2024)
[arXiv:2311.00450 [hep-ph]].

\bibitem{Miller:2007ri}
M.~L.~Miller, K.~Reygers, S.~J.~Sanders, and P.~Steinberg,
Ann. Rev. Nucl. Part. Sci. \textbf{57}, 205-243 (2007)
[arXiv:nucl-ex/0701025 [nucl-ex]].

\bibitem{dEnterria:2020dwq}
D.~d'Enterria and C.~Loizides,
Ann. Rev. Nucl. Part. Sci. \textbf{71}, 315-344 (2021)
[arXiv:2011.14909 [hep-ph]].

\bibitem{Shao:2020acd}
H.~S.~Shao,
Phys. Rev. D \textbf{101}, 
054036 (2020)
[arXiv:2001.04256 [hep-ph]].

\bibitem{Bierlich:2022pfr}
C.~Bierlich \textit{et al.}
SciPost Phys. Codeb. \textbf{2022}, 08 (2022)
[arXiv:2203.11601 [hep-ph]].

\bibitem{NNPDF:2017mvq}
R.~D.~Ball \textit{et al.} (NNPDF Collaboration),
Eur. Phys. J. C \textbf{77}, 663 (2017)
[arXiv:1706.00428 [hep-ph]].

\bibitem{AbdulKhalek:2022fyi}
R.~Abdul Khalek, R.~Gauld, T.~Giani, E.~R.~Nocera, T.~R.~Rabemananjara, and J.~Rojo,
Eur. Phys. J. C \textbf{82}, 
507 (2022)
[arXiv:2201.12363 [hep-ph]].

\bibitem{Bierlich:2018xfw}
C.~Bierlich, G.~Gustafson, L.~L{\"o}nnblad, and H.~Shah,
JHEP \textbf{10}, 134 (2018)
[arXiv:1806.10820 [hep-ph]].

\bibitem{Ceccopieri:2025edn}
F.~A.~Ceccopieri, F.~Fornetti, E.~Pace, M.~Rinaldi, G.~Salm{\`e} and N.~Iles,
Eur. Phys. J. C \textbf{85}, 
1265 (2025)
[arXiv:2507.02495 [nucl-th]].

\bibitem{Fedkevych:2025lgp}
O.~Fedkevych, J.~R.~Gaunt and S.~Smith,
JHEP \textbf{04}, 166 (2026)
[arXiv:2510.04554 [hep-ph]].

\bibitem{Gribov:1973jg}
V.~N.~Gribov,
[arXiv:hep-ph/0006158 [hep-ph]].

\bibitem{Brodsky:1994kf}
S.~J.~Brodsky, L.~Frankfurt, J.~F.~Gunion, A.~H.~Mueller and M.~Strikman,
Phys. Rev. D \textbf{50}, 3134-3144 (1994)
[arXiv:hep-ph/9402283 [hep-ph]].

\bibitem{Domdey:2009bg}
S.~Domdey, H.~J.~Pirner and U.~A.~Wiedemann,
Eur. Phys. J. C \textbf{65}, 153-162 (2010)
[arXiv:0906.4335 [hep-ph]].

\bibitem{Frankfurt:2010ea}
L.~Frankfurt, M.~Strikman and C.~Weiss,
Phys. Rev. D \textbf{83}, 054012 (2011)
[arXiv:1009.2559 [hep-ph]].

\bibitem{Corke:2011yy}
R.~Corke and T.~Sjostrand,
JHEP \textbf{05}, 009 (2011)
[arXiv:1101.5953 [hep-ph]].

\bibitem{Woods:1954zz}
R.~D.~Woods and D.~S.~Saxon,
Phys. Rev. \textbf{95}, 577-578 (1954).

\bibitem{DeVries:1987atn}
H.~De Vries, C.~W.~De Jager, and C.~De Vries,
Atom. Data Nucl. Data Tabl. \textbf{36}, 495-536 (1987).

\bibitem{Frankfurt:2011cs}
L.~Frankfurt, V.~Guzey and M.~Strikman,
Phys. Rept. \textbf{512}, 255-393 (2012)
[arXiv:1106.2091 [hep-ph]].

\bibitem{Helenius:2012wd}
I.~Helenius, K.~J.~Eskola, H.~Honkanen and C.~A.~Salgado,
JHEP \textbf{07}, 073 (2012)
[arXiv:1205.5359 [hep-ph]].

\bibitem{Kopeliovich:2022jwe}
B.~Z.~Kopeliovich, M.~Krelina, J.~Nemchik and I.~K.~Potashnikova,
Phys. Rev. D \textbf{105}, no.5, 054023 (2022)
[arXiv:2201.13021 [hep-ph]].

\bibitem{James:2004xla}
F.~James and M.~Winkler,
\textit{MINUIT User's Guide} (CERN, Geneva, 2004).

\bibitem{Eskola:2021nhw}
K.~J.~Eskola, P.~Paakkinen, H.~Paukkunen, and C.~A.~Salgado,
Eur. Phys. J. C \textbf{82}, no.5, 413 (2022)
[arXiv:2112.12462 [hep-ph]].

\end{thebibliography}
\end{document}